\documentclass[%
 prl,
 amsmath,amssymb,
 reprint,%
]{revtex4-1}

\usepackage{graphicx}
\usepackage{dcolumn}
\usepackage{bm}

\usepackage[utf8]{inputenc}
\usepackage[T1]{fontenc}
\usepackage{mathptmx}
\usepackage{etoolbox}

\makeatletter
\def\@email#1#2{%
 \endgroup
 \patchcmd{\titleblock@produce}
  {\frontmatter@RRAPformat}
  {\frontmatter@RRAPformat{\produce@RRAP{*#1\href{mailto:#2}{#2}}}\frontmatter@RRAPformat}
  {}{}
}%
\makeatother
\begin{document}

\preprint{AIP/123-QED}

\title[ Unusual phase coexistence ...]{ Unusual phase coexistence regime across the ferroelectric–paraelectric transition}

\author{Miloš Kopecký}
\author{Jiří Kub}
\author{Esther de Prado}
\author{Jiří Hlinka}
\thanks{Corresponding author. Email: hlinka@fzu.cz}
\affiliation{FZU - Institute of Physics of the Czech Academy of Sciences\\%
Na Slovance 2, 182 00 Prague 8, Czech Republic}


\begin{abstract}
A PbTiO$_{3}$ epitaxial thin film  grown on a DyScO$_{3}$ substrate, was studied by x-ray diffraction in the temperature range of 20$^\circ$C - 500$^\circ$C.
It is found that the ferroelectric to paraelectric phase transition proceeds through a peculiar transitional coexistence state extending between $T_- \approx 402^{\circ}{\rm C}$ and $T_+ \approx 414^{\circ}{\rm C}$. 
Within this temperature interval, at a temperature of $T_{3/2} \approx 405^{\circ}{\rm C}$, a domain structure transition occurs, at which the three-domain $c/a_1/a_2$ polytwin phase transforms into an ordinary $c/a_2$  twinned phase.
It is proposed that the  coexistence of two independent twinned phases with the paraelectric phase is enabled by vertical strain gradients in the film.
\end{abstract}

\maketitle


Ferroelectric films enable to miniaturize optoelectronic and electromechanical devices \cite{m1,m2,m3,m4,m5}. 
Their dielectric, piezoelectric, and optical properties are related to their domain structure. 
The latter is  affected by various aspects of the thin film growth methodology, but the key element is the mechanical stress arising from the lattice mismatch at the film/substrate interface \cite{m6,r1,r2, Streiffer02,Kwak94,Speck95,Theis1997}.

PbTiO$_{3}$ is a typical representative of  ferroelectric and ferroelastic perovskite oxide material.
In its bulk stress-free form, it undergoes the cubic to tetragonal phase transition  at about 490$^\circ$C.
Epitaxial films  grown on [001]-oriented cubic or pseudocubic substrates are typically polarized perpendicularly to the surface, or are in  a ferroelastically twinned state involving two or more 
of the three ferroelastic states of stress-free PbTiO$_{3}$.
Consequently, most of the observed textures can be understood as being composed of $a_1$, $a_2$, or $c$ ferroelastic domains, specifying whether the unique tetragonal axis is along the   [100],
[010], or  [001] directions of the pseudo-cubic parent phase, respectively.
The ferroelastic nanodomains are usually separated by  boundaries with approximately $\{ 110 \}$-type orientations\cite{m6,Theis1997}, as in the so-called permissible domain walls\cite{Marton} of stress-free PbTiO$_{3}$.

Since electrostriction-driven spontaneous strain of PbTiO$_{3}$ causes elongation along the polar axis and contraction in the perpendicular directions, a large compressive epitaxial lattice mismatch favors the $c$-domain state
while a large tensile epitaxial lattice mismatch favors the $a_1/a_2$ twinned state. 
In the intermediate region, a competition or symbiosis of the  $a_1/a_2$ and $c/a$ lamellar twins  can be expected.
In fact, when the effective in-plane lattice parameters
of the substrate fall between the  $c$ and $a$ lattice parameters of the stress-free PbTiO$_{3}$, the average lattice parameter of the three-domain polytwin $c/a_1/a_2$ with appropriately chosen domain fractions can  always match that of the substrate \cite{m6}.
These $c/a_1/a_2$ adaptive polytwin structures are systematically observed in relatively thick films, where the energetically costly topological defects at which two or more ferroelastic domains merge can be accommodated \cite{m6, Royt2001, Alpa1998, tova23, tova25, tova26,m10, m11, m12, m13, Feigl2014, Feigl2015, Simon,Sato2022, Highland2014,Theis1997,catalan}.

\begin{figure}[ht]
\centering
\includegraphics[width=.69\columnwidth]{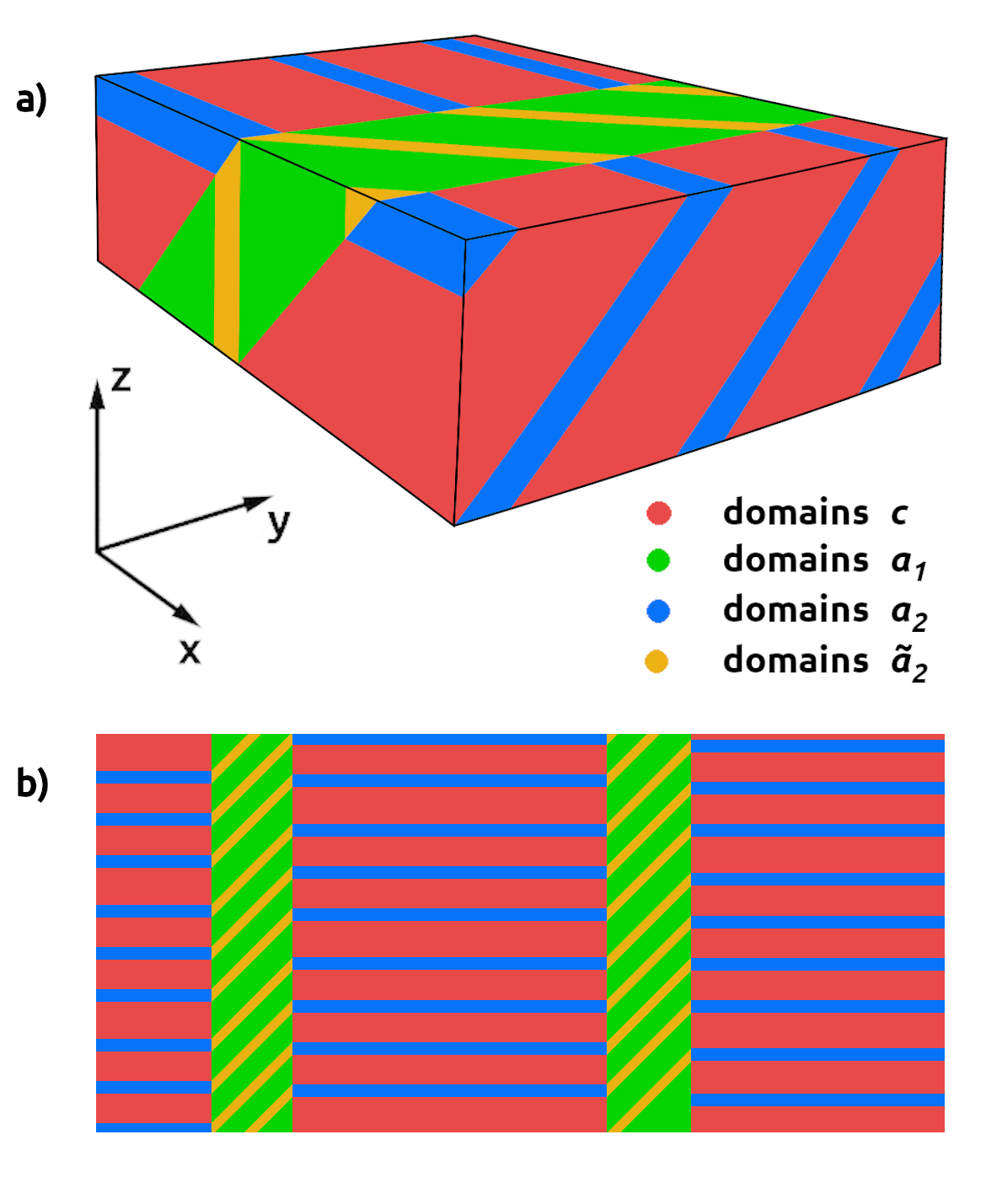}
\caption{Type-III (Swedish ladder) $c/a_1/a_2$ polytwin texture encountered in epitaxial lead titanate films. 
(a) Perspective view at the junction of the $c/a_2$ (red/blue) and $a_1/\tilde{a}_2$ (green/yellow) lamellae and (b) top view of a larger portion of the film.
Cartesian $x$, $y$, and $z$ axes are along $\bf{a}[\rm s]$, $\bf{b}[\rm s]$ and $\bf{c}[\rm s]$ pseudo-cubic lattice vectors of the substrate.
}
\label{fig:o1}
\end{figure}


The nanodomain junctions encountered in the $c/a_1/a_2$ polytwin with the type-III (Swedish ladder)  architecture \cite{ m10, Swedish}, which is the subject of the present work, are shown in Fig.\,\ref{fig:o1}a. 
The prevailing $c$-domain state (in red) is intersected by $a$-domains of three types: the $a_1$-domains (in green) cause the enhancement of the average lattice parameter along the [100]-direction, while $a_2$ (blue) and $\tilde{a}_2$ (yellow) domains are enhancing the average lattice parameter along the [010]-direction.
The  $a_2$  and $\tilde{a}_2$ domains have antiparallel spontaneous polarization. 
Moreover, the $a_2$ domains are more elongated than the $\tilde{a}_2$ domains\cite{m10}.
Consequently, in a larger scale top view of the film (Fig.\,\ref{fig:o1}b), the $a$-domain pattern reminds of a Swedish ladder  with its side rails formed by the thicker and more distant $a_1/\tilde{a}_2$ strips and its  horizontal crossbars formed by the dense set of the narrow and quite regularly spaced  $a_2$ domain strips\cite{m10,Swedish,tova23}.

For (001)-oriented tetragonal substrates such as SrTiO$_{3}$, the $x$ and $y$ in-plane directions are equivalent.
This is not the case for orthorhombic substrates such as (110)-DyScO$_{3}$ or (110)-TbScO$_{3}$. 
Previous experiments have revealed that the narrow $a$-domains are systematically aligned\cite{Swedish,m10,tova23, Simon}  along the shorter in-plane pseudo-cubic lattice parameter of the substrate, $a_0[\rm s]=|{\bf a}[\rm s]|$, as set in Fig.\,\ref{fig:o1}a.

In this work, we investigated the ferroelectric-to-paraelectric phase transformation in a 400-nm thick PbTiO$_{3}$ film grown on (110)-DyScO$_{3}$ by means of synchrotron x-ray diffraction. 
Ferroelectric and paraelectric phases were found to coexist over a remarkably broad temperature interval of about 402-414\,$^\circ$C. 
Within this temperature interval, at a temperature $T_{3/2} \approx 405$\,$^\circ$C,  an abrupt phase transition is found, at which the 3-state Swedish ladder domain structure $(c/ a_2)$/$(a_1/\tilde{a}_2)$  is converted into a simply twinned 2-state $c/ a_2$  structure. The significance of these results is discussed.

{\it Ambient temperature domain structure.}
Similarly as in twinned ferroelectric crystals, the mechanical compatibility of the adjacent domains in the film causes slight tilts of their lattices with respect to the crystallographic axes of the substrate.
Changes in lattice parameters and these lattice tilts result in split pseudocubic Bragg reflections.
Each domain state can be assigned to its own set of Bragg reflections,
as discussed, e.g., in Ref.\,\onlinecite{m12}.
Moreover, the orthorhombic symmetry of the (110)-DyScO$_{3}$ substrate implies that along with the ferroelastic pattern charted in Fig.\,\ref{fig:o1}a,  there must be three other energetically equivalent variants\cite{m11}, 
which can be generated by mirroring the configuration of Fig.\,\ref{fig:o1}a via the (100) and/or (010) planes. 
As a result, each Bragg reflection is actually split into 16 individual reflections.
 Reciprocal space maps  as obtained in the present experiment at room temperature are shown  in the End Matter. 
 
The measurement reveals that the crystal lattice of the film differs from that of the ambient stress-free PbTiO$_{3}$.
First, the two smaller lattice parameters are not equal in the film.
Whether inspecting the state $c$, $a_1$ or $a_2$, one of its two smaller lattice parameters is smaller than the other by at least 0.1\%. 
Secondly, the largest lattice parameters are by more than 0.1\% smaller than the 4.152$\,\text{\AA}$ value of the stress-free material\cite{tova25, tova23, Shirane1956}.
Moreover, different domain states differ in their structure.
In particular, 
 the longest lattice parameters of ferroelastic domains of $a_1/\tilde{a}_2$ strip are about 4.14$\,\text{\AA}$, while  4.11$\,\text{\AA}$ is found in domains of the $c/a_2$ strip.
In the following, the lattice parameter  measured along the Cartesian direction $x,y$ and $z$ of the substrate is labeled $a_0,b_0$, and $c_0$, respectively, and the domain state is given in square brackets, such as, {\it e.g.} $a_0[a_2]$ or $b_0[c]$.

\begin{figure}[ht]
\centering
\includegraphics[width=.89\columnwidth]{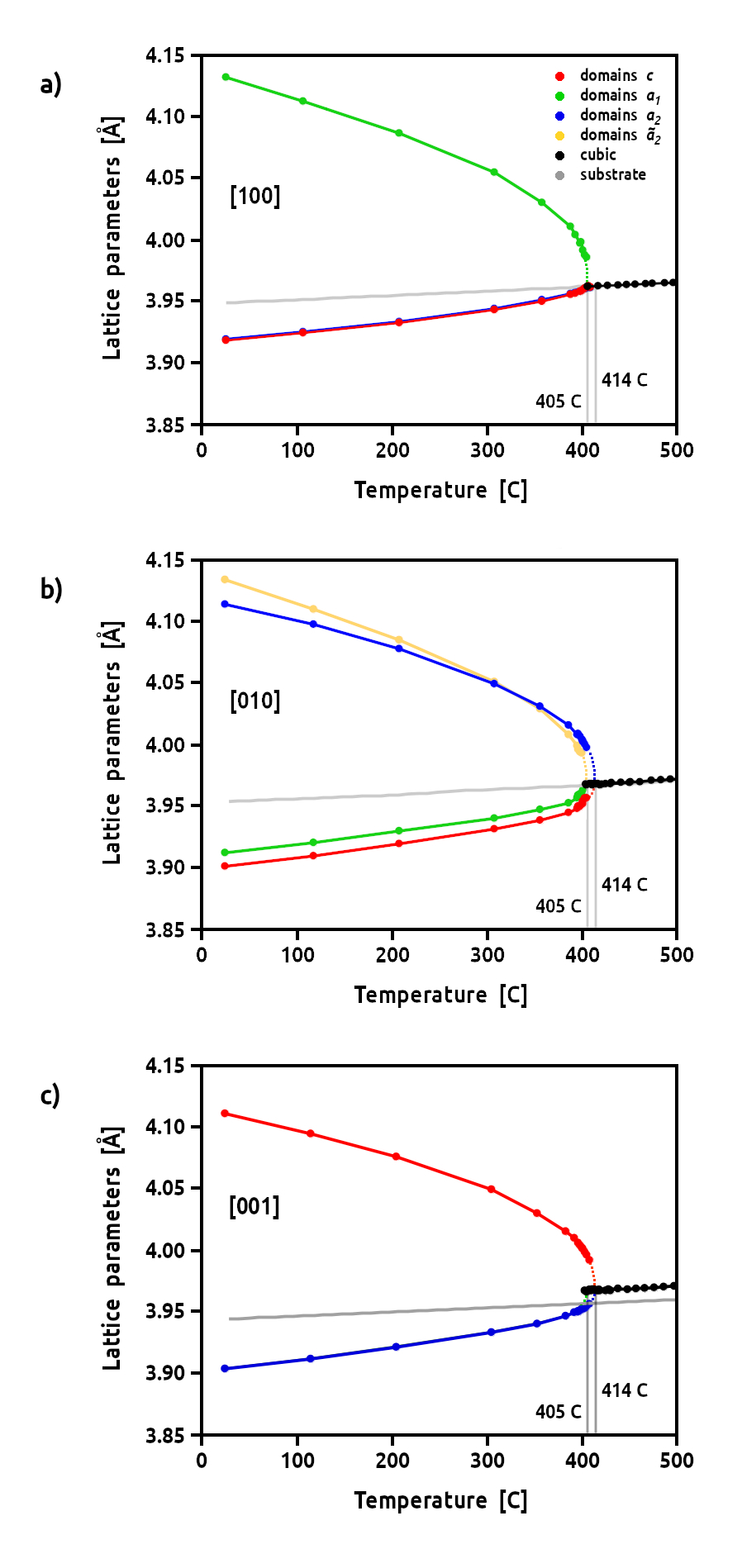}
\caption{Temperature dependence of lattice parameters of the investigated PbTiO$_{3}$ film (on heating). The in-plane lattice parameters along [100] (a), along [010] (b) and the out of plane lattice parameters along [001] (c)  directions  are plotted together with the pseudocubic lattice parameters of the substrate.
}
\label{fig:o3}
\end{figure}

{ \it Temperature dependence of lattice parameters.}
Next, the positions of the split Bragg reflections were tracked from room temperature up to about 500\,$^\circ$C. 
The most interesting outcome of the measurements is the temperature dependence of the lattice parameters 
 in each type of the ferroelastic domain state.
Temperature-dependent lattice parameters 
are shown in Fig.\,\ref{fig:o3}.

Several characteristic temperatures were identified.
First, the measurement reveals 
that below $T_{\rm cross}=$330$^\circ$C, the $c$ lattice parameters in $a_1/\tilde{a}_2$ strips are greater than those in $c/a_2$ strips, while the opposite is true above this temperature (see Fig.\,\ref{fig:o3}b).
Second,  at the temperature of $T_{-}=$ 402$^\circ$C, another  diffraction spot appears in reciprocal space maps in addition to the split reflections from the ferroelastic domains.
This can be regarded as the onset of the transformation toward the paraelectric phase, since the reflection persists up to the highest measured temperatures after all other film reflections have vanished.
The third characteristic temperature, $T_{3/2}=$405$^\circ$C, is the temperature at which
the   $a_0[c]$ and $a_0[a_2]$ lattice parameters   meet the substrate value $a_0[\rm s]$ (see Fig.\,\ref{fig:o3}a).
It can be associated with the $(c/ a_2)$/$(a_1/\tilde{a}_2)$ to $c/ a_2$ conversion,  as it also corresponds to the temperature at which the intensity of the reflections from the $a_1/\tilde{a}_2$ strips vanishes.
Therefore, $T_{3/2}$ represents a triple point, where $c/ a_2$, $a_1/\tilde{a}_2$ and the  paraelectric phases coexist.
Finally, at about $T_{+}=$\,414$^\circ$C, the intensity of all split reflections associated with ferroelastic domains vanishes.
We emphasize that this temperature is more than 70$^\circ$C {\it below} the known ferroelectric phase transition of the stress-free PbTiO$_{3}$.

\begin{figure}[ht]
\centering
\includegraphics[width=\columnwidth]{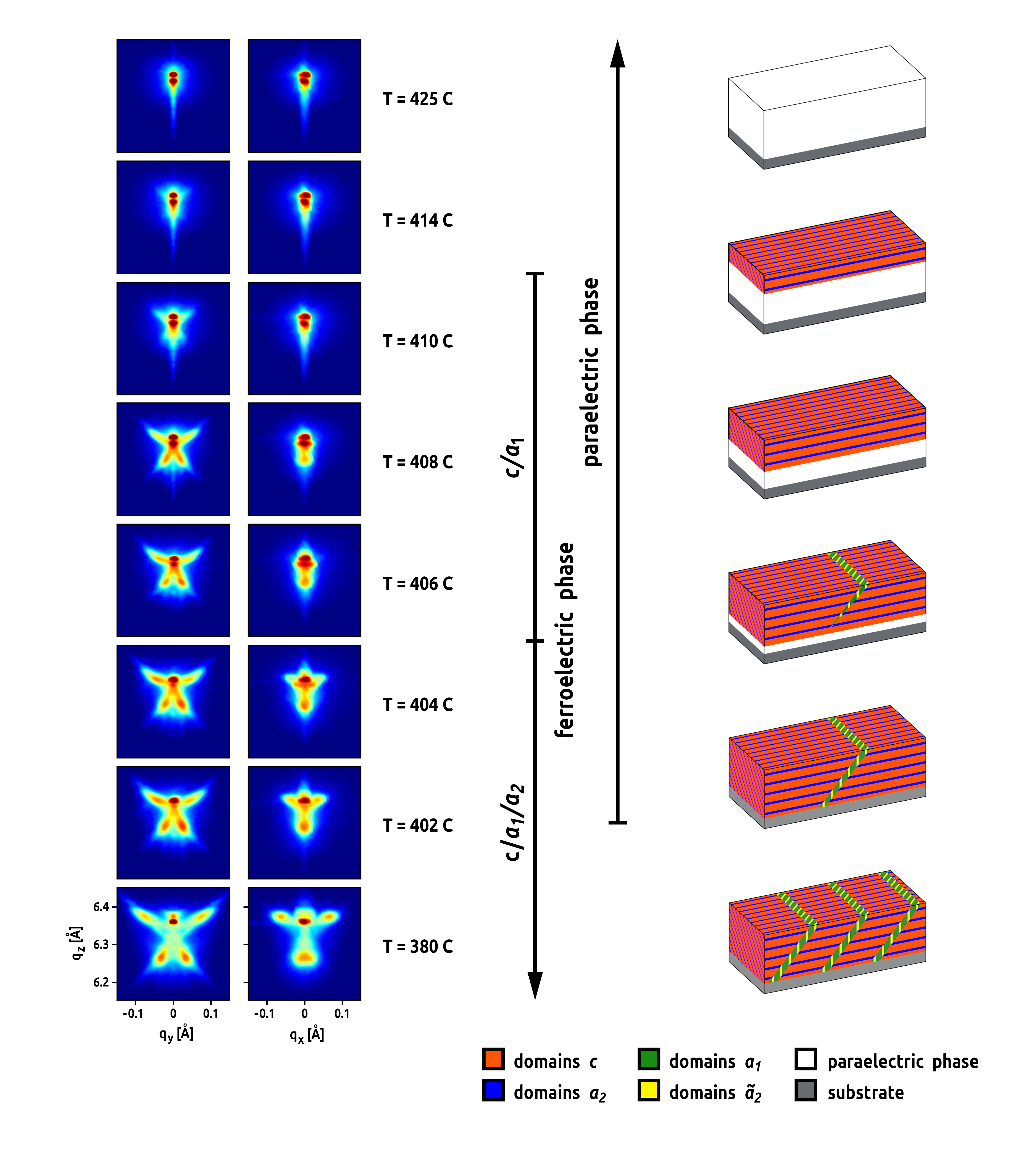}
\caption{Reciprocal space maps in the vicinity of the (004) reflection along $x$, and $y$ axes  at several temperatures close to the phase transition temperature (left). The inferred modification of the domain structure in the film during the phase transition from the ferroelectric phase to the paraelectric phase is drawn schematically (right). 
}
\label{fig:o4}
\end{figure}



 { \it Reduced critical temperature and spontaneous strain.}
 Within the standard Ginzburg-Landau-Devonshire theory for perovskite oxide ferroelectrics, the spontaneous strain is a quadratic form of the spontaneous polarization, and both order parameters are directly linked by electrostriction\cite{Marton}.
 If the origin of the reduced spontaneous strain was in the  temperature-dependent coefficient of the leading quadratic term of the Landau expansion, the ambient-temperature tetragonality $\tau=(c/a)-1$ would be proportional to the temperature interval between the critical temperature and the ambient temperature. 
 Comparing the 6\% ambient tetragonality\cite{Shirane1956} of the stress-free PbTiO$_{3}$ with the observed, 5\%-ambient tetragonality in the present film, the expected 17-20\%-reduction of the temperature interval would imply a critical temperature in the film at about $  (20 + (490-20)/1.2)^\circ{\rm C}\approx 410^\circ {\rm C}$, suggesting that both strain and temperature shifts might be indeed correlated.
 
 Nevertheless, our measurements also show that the phase transition in the film is significantly closer to the tricritical point than in the stress-free material, so that the higher-order terms in the Landau expansion are surely renormalized in the film, too.
Indeed, while the $c_0[c]$ lattice parameter of the majority domain state $c$ shows a stepwise change near the ferroelectric phase transition, indicating that its first-order nature is maintained, in fact, the step is much smaller than in the stress-free PbTiO$_{3}$. 
The effective tetragonality  in $c$-domain $\tau_c =2c_0[c]/(a_0[c]+b_0[c])-1$ at $T_+$ is clearly less than 1\%, compared to about 2\% reported\cite{Shirane1951} for the stress-free PbTiO$_{3}$.
This suggests a 3-4 times lower
absolute value of the quartic term coefficient in the effective Landau free energy potential of the film compared to stress-free PbTiO$_{3}$, which can be ascribed to electrostrictive renormalization due to mechanical clamping effects\cite{m18,m19,m20}. 

Interestingly, the observed $T_+$ is quite close to the paraelectric Curie-Weiss temperature $T_0$ of the stress-free PbTiO$_{3}$\cite{Shirane1951,Samara1971,HRSPTO}.
However, the electrostrictive renormalization of the quartic term coefficient also implies a Clausius-Clapeyron-like downshift of the quadratic term coefficient in the Landau expansion, and this should result in an overall upshift of the phase transition temperature.
Therefore, we infer that there are other mechanisms causing the upshift of the quadratic term coefficient, such as inefficient screening of the 
depolarization field film-substrate interface\cite{Highland2014,Lichtensteiger2005}.

{\it Heterophase character of the Swedish ladder polytwin.}
 The $c/a_2$ structure  is known to have a quasi-regular lamellar  arrangement 
 with a periodicity smaller than the thickness of the PbTiO$_3$ film \cite{m12,m15,tova23,tova25,tova26}.
 Therefore, the largest facets of the $c$ and $a_2$ domains are their common interfaces, the 90$^\circ$-ferroelectric walls. 
 The continuity of the normal components of the  polarization and of the tangential components of the strain near these interfaces favors equal magnitudes of the spontaneous polarization and strain in the adjacent domains, respectively. 
 It results in relationships $c_0[c] \approx b_0[a_2]$, $b_0[c] \approx c_0[a_2]$, and $a_0[c] \approx a_0[a_2]$, verified in our experiment.
 In other words, the high density of ferroelastic domain walls in the $c/a_2$ twinned strip strongly links  the structures of the $c$ and $a_2$ domains, and these links justify the approximation in which the whole $c/a_2$ strip is itself considered as a secondary (super)domain of the   $c/a$-twinned "phase". 
 
 To a large extent, the same applies to the $(a_1/\tilde{a}_2)$ strips, so that $a_0[a_1]\approx b_0[\tilde{a}_2], b_0[a_1]\approx a_0[\tilde{a}_2]$ and $ c_0[a_1]\approx c_0[\tilde{a}_2]$. However, the overall surface area and density of  the secondary domain walls between the $c/a_2$ and  $(a_1/\tilde{a}_2)$ superdomains is much lower, and
 so the resulting coupling between the order parameters is much weaker. 
 In other words, the structures of $c/a_2$ and $(a_1/\tilde{a}_2)$ strips are mutually more independent, and  thus each responds as a domain of a different phase. 
 This weak coupling between the $c/a_2$ and $(a_1/\tilde{a}_2)$ domains is revealed in the measured differences between  $b_0[\tilde{a}_2]$ and $b_0[a_2]$ or between $b_0[c]$ and $b_0[a_1]$ parameters shown in Fig.\,\ref{fig:o3}b.
Similarly, lattice parameters of the  $a_1/\tilde{a}_2$ strip tend to approach paraelectric ones slightly above the $T_{3/2}$ while lattice parameters of the $c/a_2$ strip tend to approach the paraelectric lattice parameters slightly above $T_{+}$.
In this sense, the present measurements demonstrate that the Swedish ladder domain texture has a character of a { \em heterophase} polytwin\cite{m20}.

 {\it Nanodomain structure transformation.}
Nanodomain pattern transformations in epitaxial  thin films of tetragonal ferroelastics has been addressed by phenomenological theory \cite{m19,m20,Royt2001, Alpa1998,Speck95, Feigl2014,Feigl2015}. 
The most relevant to the present work is the prediction of the transformation from the three-domain polytwin state to the $c$ single-domain ferroelastic state\cite{Royt2001}. 
Since 
this theory was aimed at (001) cubic substrates, there is only one in-plane lattice parameter of the substrate in the theory. 
The main idea is that the three-domain polytwin allows to relax the macroscopic strain as long as the effective substrate lattice parameter falls between the relaxed lattice parameters of the tetragonal ferroelastic material of the film. 
Consequently, the $T_{3/1}$ phase transition should occur when the misfit strain between the substrate and the relaxed $c$-domain is zero. \cite{Royt2001, Alpa1998} 

Here, the in-plane lattice parameters of DyScO$_{3}$ are unequal.
Thus, the zero misfit condition for $a_0[\rm s]$ should be met at a  somewhat lower temperature\cite{Feigl2014,Feigl2015} than that for $b_0[\rm s]$.
In fact, we observe (see Fig.\,\ref{fig:o3}a) that $a_0[c] \approx a_0[a_2]$ approaches $a_0[\rm s]$ as $T_{3/2}$ is reached from below and the zero $x$-misfit  $a_0[c] = a_0[a_2]= a_0[\rm s]$ is maintained in the 2-state phase above $T_{3/2}$.
Similarly $b_0[c]$ approaches $b_0[\rm s]$ as $T_{+}$ is reached from below (see Fig.\,\ref{fig:o3}b), and the zero $y$-misfit condition $b_0[c] =b_0[a_2] = b_0[\rm s]$ is likely the onset condition of the $T_{+}$ transition itself.

 { \it Phase coexistence due to vertical strain gradients.}
The most unexpected result is  the broad temperature range of the coexistence of the ferroelastic and paraelastic phases, within which we observed a natural evolution of the order parameter magnitudes
and even an additional phase transition.
Considering the slowness of the measurement protocol and the negligible thickness of the film, neither the long transformation kinetics nor the residual temperature gradients appear to be likely reasons for such  coexistence. 
Rather, based on the measurements of the stress-free lead titanate, thermal hysteresis smaller than  1-2$^\circ$C is expected\cite{Sun1993}.

The phase transition in lead titanate crystals is known to proceed by propagation of phase fronts separating the paraelectric and twinned ferroelectric regions\cite{Dec}.
The phase front tends to be a planar interface with a particular crystallographic orientation that ensures the compatibility of the macroscopic strain on both sides of the phase front (the so-called invariant strain plane or habit plane)\cite{DiDomenico,Bednyakov}.

In the case of epitaxial films, both phases are subject to the same epitaxial constraint.
Therefore, the average mechanical compatibility is always satisfied in planes parallel to the substrate surface.
The  vertical phase fronts are probably also energetically affordable, especially in ultrathin films, where the thicness-integrated lattice mismatch between the ferroelectric and paraelectric out-of-plane lattice parameters yields less than
an elementary unit cell difference.
However, the lateral motion of such interfaces would probably be significantly hindered by vertical mismatch between the paralelectric phase and $c$-domain islands so that the motion of such phase-fronts would be less efficient in realization of the phase transformation.

We therefore speculate that in the phase coexistence range $T_-<T<T_+$,
the paraelectric phase front is parallel to the substrate, as sketched in Fig.\,\ref{fig:o4}.
The position of the phase front within the height of the film represents an additional degree of freedom of the system.
Since the electric and mechanical constraints imposing the overall downshift of the critical temperature act primarily near the film-substrate interface, we assume that the phase front moves from the substrate towards the film surface as the temperature increases from
$T_-$ to $T_+$.

In fact, the broadening of the split Bragg reflections in reciprocal space maps (see Fig.\,\ref{fig:o3} and \ref{fig:o4}) suggests that there is some spread of the lattice parameters in each domain state, possibly associated with order parameter gradients across the thickness of the film\cite{Everhardt2020}. 
This is expected to be caused by the vertical gradient of strain relaxation, reflecting the natural asymmetry between the mechanically clamping and insulating substrate interface and the mechanically-free top surface with possibly better passivated bound charge. 
Such an effective built-in  external field gradient is known to break up the global thermodynamic equilibrium and the usual Gibbs phase rule. 
Similarly to 
phase changes in materials subject to a gravitational field or frozen composition gradients,
this built-in gradient can explain the expansion of the  phase transition into a finite temperature region of the phase coexistence.
Some of our  findings and arguments  possibly apply to the similarly peculiar behavior of BaTiO$_{3}$ films grown on NdScO$_{3}$ substrates, where similar effects due to the vertical built-in  gradients were considered previously \cite{Everhardt2020}.

In summary, detailed synchrotron x-ray diffraction investigations allowed us to reveal and clarify several interesting aspects of the ferroelectric to paraelectric phase transition in a thin film grown on a DyScO$_{3}$ substrate.
Our experiment revealed that the ambient-strain relaxation reaches only about 80\% of the stress-free values, the phase transition temperature is reduced by more than 70$^{\circ}{\rm C}$,
and the transitional jump of the spontaneous strain is reduced compared to that of the stress-free PbTiO$_{3}$.
Inspection of lattice parameters in individual ferroelectric domain states demonstrates that $c/a_2$ and $a_1/a_2$ superdomain strips have independent temperature dependence,
allowing the Swedish ladder texture to be identified as a heterophase polytwin.
The  phase transition between the $c/a_1/a_2$ Swedish ladder texture and an ordinary $c/a_2$ twin is associated with the split zero misfit condition for the $c$ domain 
on a substrate with slightly different in-plane lattice parameters.
This phase transition happened to occur while the ferroelectric part of the film is already in  coexistence with the paraelectric phase.
This peculiar situation is related to the unusually extended temperature region of the paraelectric and ferroelectric phase coexistence.
It is argued that the formation of the wide transitional region results from the  vertical gradients of the built-in strain or field.
Similar effects may be encountered and exploited in other ferroelectric epitaxial thin films and possibly can be used to design desired properties like a wide temperature range of high susceptibilities.

\newpage
\appendix
\onecolumngrid
\section*{End Matter}
\twocolumngrid
\subsection{Experimental details}

The  PbTiO$_{3}$ film used in this study was grown on a (110)-DyScO$_{3}$ substrate by pulsed laser deposition at around 600\,$^\circ$C to avoid formation of interfacial dislocations during growth, similarly as in the case of the films reported in Refs.\,\onlinecite{Perantie,Dejneka}.
The original 10$\times$10\,mm$^2$  sample was cut into several pieces of 1$\times$1.2\,mm$^2$ dimensions.
These smaller samples were attached to the tip of a quartz capillary using a temperature resistant glue.
The capillary was mounted on a goniometer head in a standard manner. 
The  experiments were carried out on the x-ray diffraction beamline (XRD1) at Sincrotrone Trieste, Italy. 
The energy of the incident beam was set to 17.712\,keV corresponding to the wavelength of 0.7$\,\text{\AA}$. 
The size of the incident beam was defined by a 100-micron-diameter pinhole positioned in front of the sample.
A Pilatus 2M large-area  detector  was placed at a distance between 260\,mm behind the sample. 
The beam energy and the geometrical parameters of the experiment were refined using one of the samples sprinkled with a LaB$_6$ powder calibrant.  
%

\begin{figure}[ht]
\centering
\includegraphics[width=.99\columnwidth]{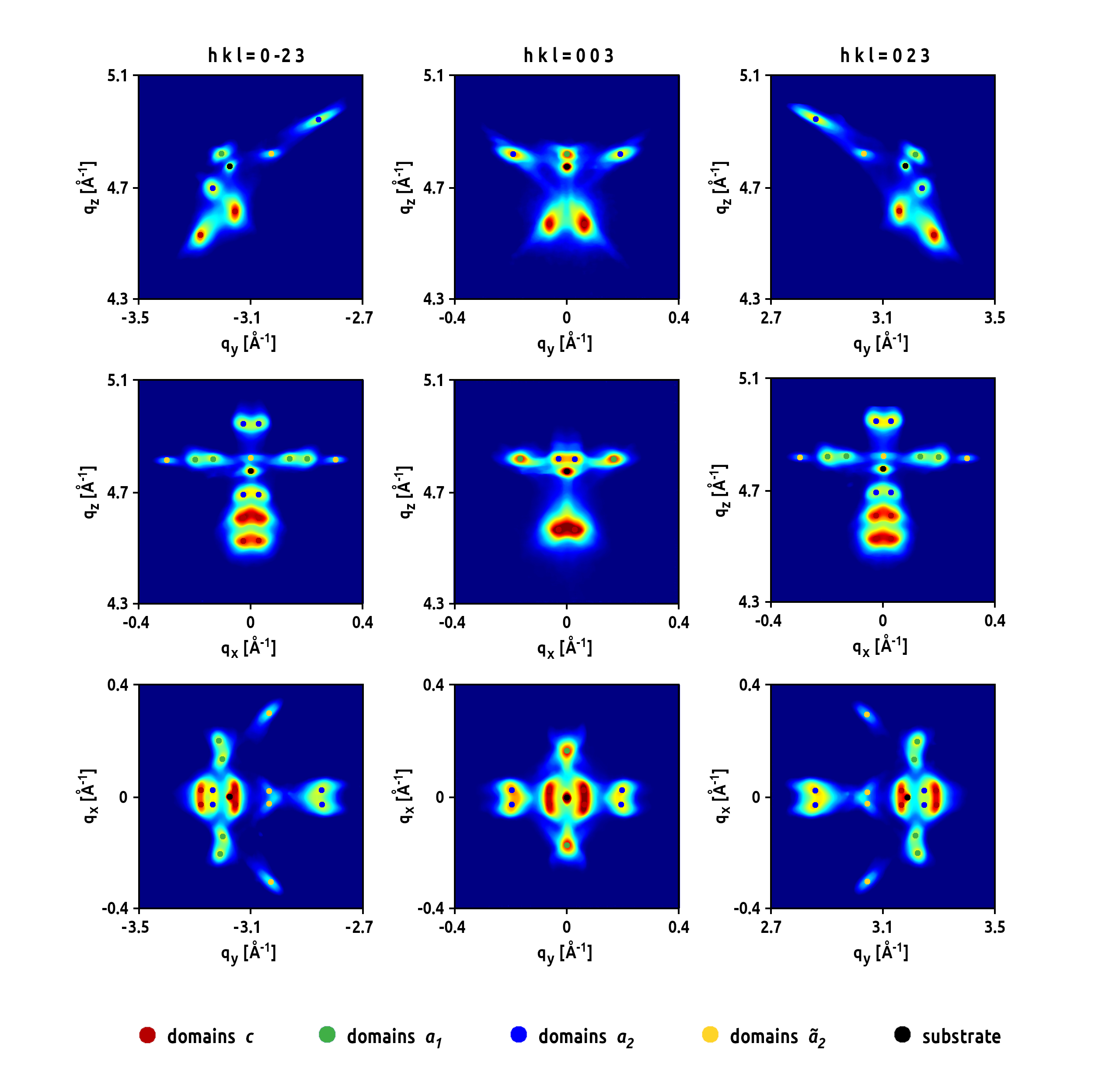}
\caption{ Room-temperature reciprocal-space maps of the 400\,nm-thick PbTiO$_{3}$ film grown on the DyScO$_{3}$ substrate. Panels show the vicinity of the $0\bar{2}3$, 003, and 023 Bragg reflections (rows). Rows correspond to the views along $x$, $y$, and $z$ axes for each reflection. Diffraction maxima corresponding to individual domain types and to the substrate are marked by small color and black circles, respectively.  
}
\label{fig:o2}
\end{figure}


The sample temperature was controlled by an XDS Oxford gas blower that circulated hot air on the illuminated sample surface. 
The sample was heated with variable temperature steps starting with 100$^\circ$C at lower temperatures and about 2$^\circ$C near the phase transition.
The temperature dependence of the out-of-plane lattice parameters (the direction [001])  in the film was determined from the positions of the  (004) split diffraction spot, whereas the in-plane lattice parameters in the direction [010] were determined using the distance of (0$\bar{2}$3) and (023) reflections in reciprocal space. 
	Analogically, the in plane lattice parameters in the crystallographic directions [100] were obtained from diffraction spots ($\bar{2}$03) and (203). Another sample attached to the capillary in different geometry (rotated 90$^\circ$ around the surface normal with respect to the previous one) was used for this purpose. The crystallographic axis [010] of the lead titanate film was oriented close to the plane of incidence in this case; the diffractometer axis $\varphi$ was very close to the [100]-crystallographic axis. The three dimensional part of the reciprocal space, including reflections  ($\bar{2}$03), (003), (203), and (004), was mapped using the same data collection parameters as in the case of the first sample.

 Reciprocal space maps around reflections (0$\bar{2}$3), (003), and (023) from directions [100], [010], and [001] as obtained in the present experiment at room temperature are shown  in  Fig.\,\ref{fig:o2}.
The assigned correspondence of the individual diffraction spots with each domain state in the film is indicated by small color markers.

\subsection{Acknowledgments}
\begin{acknowledgments}
This work has been supported by the project NEPHEWS under the Grant Agreement No 101131414 within EU Framework Programme for Research and Innovation Horizon Europe and by the Ferroic Multifunctionalities project, supported by the Ministry of Education, Youth, and Sports of the Czech Republic, Project No. CZ.02.01.01/00/22\underline{\,\,\,}008/0004591, co-funded by the European Union. The authors thank Elettra Sincrotrone Trieste for providing access to its synchrotron radiation facilities and their beamline experts, Maurizio Polentarutti and Giorgio Bais, for the assistance in using X-ray diffraction beamline. We also acknowledge our colleagues P. Bérešová, F. Borodavka,  P. Ondrejkovič, M. Pa\'{s}ciak,  K. Tesař,   and  M. Tyunina from the Institute of Physics, Czech Acad. Sci. for useful discussions and the experimental characterization prior to these investigations, in particular Dr. P. Ondrejkovič for critical reading of the manuscript, and Prof. M. Tyunina for providing us with the high quality thin film sample.
\end{acknowledgments}

\end{document}